\documentclass[useAMS,usenatbib]{mn2e}
\usepackage{graphicx}
\usepackage{amsmath}

\title[Black Hole Shadow Image and Visibility Analysis of Sgr~A*]
{Black Hole Shadow Image and Visibility Analysis of
Sagittarius~A*}
\author[L. Huang et al.]{ Lei Huang$^{1,3}$,
Mike Cai$^{4}$, Zhi-Qiang Shen$^{1,2}$, and Feng Yuan$^{1,2}$\\
$^{1}$Shanghai Astronomical Observatory, Chinese Academy of Sciences, Shanghai 200030, China\\
$^{2}$Joint Institute for Galaxy and Cosmology (JOINGC) of ShAO and USTC, Shanghai 200030, China\\
$^{3}$Graduate School of the Chinese Academy of Sciences, Beijing 100039, China\\
$^{4}$Academia Sinica, Institute of Astronomy and Astrophysics,
Taipei, China}
\begin{document}


\pagerange{\pageref{firstpage}--\pageref{lastpage}} 

\maketitle

\label{firstpage}

\begin{abstract}
The compact dark objects with very large masses residing at the
centres of galaxies are believed to be black holes. Due to the
gravitational lensing effect, they would cast a shadow larger than
their horizon size over the background, whose shape and size can
be calculated. For the supermassive black hole candidate Sgr A*,
this shadow spans an angular size of about 50 micro~arc~second,
which is under the resolution attainable with the current
astronomical instruments. Such a shadow image of Sgr~A* will be
observable at about 1 mm wavelength, considering the scatter
broadening by the interstellar medium. By simulating the black
hole shadow image of Sgr A* with the radiatively inefficient
accretion flow model, we demonstrate that analyzing the properties
of the visibility function can help us determine some parameters
of the black hole configuration, which is instructive to the
sub-millimeter VLBI observations of Sgr A* in the near future.
\end{abstract}


\section{Introduction}
There is compelling evidence that a $~4\times 10^6 M_\odot$ black
hole, associated with an extremely compact radio source
Sagittarius A* (Sgr A*), resides at the Galactic Centre
\citep*{Sd03, Gz05, Me01}. Recent Very Long Baseline
Interferometry (VLBI) observations reveal that the intrinsic size
of Sgr A* is only about 1 AU \citep{Shen05}. For such a high
degree of mass concentration, the relaxation time scale is much
shorter than the age of our Galaxy, and thus almost all the
theoretical models predict the formation of a supermassive black
hole via gravitational collapse \citep{Mz98}.

However, to conclusively prove that Sgr A* is a supermassive black
hole requires observations close to its event horizon, where
relativistic effects can not be ignored. According to General
Relativity, the photon trajectory will bend appreciably in a
strong gravitational field \citep*[see, e.g., ][]{M73, Chdr83}.
Because of this peculiar behavior, images of black holes and their
immediate neighborhood we observe are highly distorted. The true
physical structure in the very vicinity of the black hole must be
decoded from the image via a ray tracing method
\citep*[]{Ft97, Fc00, Schn04, Schn06}. In particular, if
the impact parameter is smaller than a critical value, the photon
is doomed to fall into the black hole \citep{Chdr83}. As seen by a
distant observer, this effect allows the black hole to cast a
shadow over the background source that is larger than its event
horizon.

The black hole candidate Sgr A*, at a distance of 8 kpc to us,
spans an angular size of $\sim$ 20 micro arc second ($\mu$as) in
diameter in the sky, which is beyond the resolution attainable
with the current astronomical instruments. However, the shadow
size is always of order $10GM/c^2$ in diameter, corresponding to
an angular size of $50\mu$as for Sgr~A*. In principle, this
resolution can be achieved using VLBI techniques at a baseline
length $\sim4\times10^3\rm{km}$ if observed at 1 mm wavelength
\citep*[e.g., ][]{Fc00}.

In this paper, we adopt the ray-tracing method to map the Sgr A*
black hole with the radiatively inefficient accretion flow (RIAF)
model \citep*{Y03}. We then use our simulations to estimate the
visibility function to be observed by a VLBI array operating at,
if can be designed in the future, sub-millimeter wavelengths. We
are on the verge of resolving the shadow of the presumed
supermassive black hole at the Galactic Centre.

The paper is organized as follows. Section $2$ briefly
describes the formalism we use for the ray-tracing algorithm. We
show the simulation results in Section $3$.  Then we estimate the
visibility functions in observations in Section $4$. Finally we
give a discussion in Section $5$.

Throughout this paper, we will work with geometric units where
$c=G=1$.

\section[]{Hamiltonian Geodesic Ray-tracing in Black Hole Spacetime}

In a given spacetime metric, the trajectory of a photon is
uniquely determined by the initial conditions.  To efficiently
simulate the image of a black hole and its environment, we opt to
impose our initial conditions at infinity, where the observer is,
and integrate backwards in affine parameter toward the black hole
\citep[cf.][]{RB94, Ft97}. Starting from the standard kinetic
Lagrangian, we compute the super-Hamiltonian \citep[cf.][]{M73,
FQ03} in Kerr metric but with spin parameter $a=0$, since we
discuss only the Schwarzschild black hole case in this work.
Normal variational approach yields the Hamiltonian equations of
motion in the phase space $(q^\mu,p_\mu)$, where $q^{\mu}$ and
$p_\mu$ are Boyer-Linquist coordinate and the canonical momentum
of the particle, respectively. If we choose the affine parameter
to be proper time per unit rest mass (which has a finite limit
even when the rest mass is zero in the photon case), then $p_\mu$
is interpreted as the covariant component of the four-momentum.
Similar process can be seen in, e.g., \citet{Schn04, Br06}.

As discovered by \citet{Bard72}, an infalling photon with
its impact parameter \citep{Chdr83} less than $\sqrt{27}M$ will
never escape the vicinity of the black hole.  As seen by an
outside observer, the black hole will cast a shadow of radius
$r_\text{shadow} = \sqrt{27}M \approx 5.2M$ over the background
light source. As reported by many numerical calculations (e.g.,
Falcke et al, 2000; Takahashi 2004), rotating black holes will
cast shadows of approximately the same size as well.

In the following simulation, we specify initial conditions
at a large distance (which we have chosen arbitrarily $r_0 =
200M$), and then integrate backward toward the black hole to
obtain the full geodesic.

\section{Results From Simulation}
Fig. 1 is a sketch map, defining the inclination angle $i$
\citep{Chdr83} and position angle $\Theta$ used in our
simulation. Three axes $(X,Y,Z)$ point east, north and the
direction aligning with the line of sight for an observer,
respectively. The thick blue arrow represents the direction of the
black hole's equatorial plane (the spin axis for the Kerr black
hole). The inclination angle $i\in [-\pi/2,\pi/2]$ is the angle
between the spin axis and the observer's line of sight (Z axis),
and the position angle $\Theta\in [-\pi,\pi]$ is the angle between
the projection of spin axis on $X-Y$ plane and $Y$ axis, positive
by east and negative by west. We limit $i\in [0,\pi/2]$ and
$\Theta\in [-\pi/2,\pi/2]$ in the following discussion because of
symmetry.

\subsection{Black Hole Shadow of Sgr A* with the Radiatively Inefficient Accretion Flow Model}
Even though there have been various theoretical models proposed to
explain the observations of Sgr A* \citep*[see, e.g., ][]{Me01},
the precise structure of the dominant emitting region is still
poorly known. In this paper, we only consider the RIAF model
proposed by \citet{Y03}, which can satisfy most observational
results including the observed spectrum of Sgr A* from radio to
X-ray, the flaring activity at both infrared and X-ray bands, and
its extremely faint luminosity.

In the RIAF model, the emission of Sgr A* from millimeter to
sub-millimeter is dominated by the synchrotron radiation of
thermal electrons. Thus one can determine the
frequency-dependent emissivity $j_\nu(r,0)$ in the
equatorial plane. Assuming a Gaussian distribution of
electron density in the vertical direction of a RIAF, the
emissivity distribution function is then:
\begin{eqnarray}\label{distr1}
j_\nu(r,z) = j_\nu(r,0)\exp(-\frac{z^2}{H^2}),
\end{eqnarray}
where $z$ is the vertical distance from the equatorial plane and
$H$ is the scale height determined by:
\begin{eqnarray}\label{H}
H = \frac{\alpha_s}{\Omega_\rmn{K}},
\end{eqnarray}
where $\alpha_s$ is the thermal sound speed and $\Omega_\rmn{K}$
is the Keplerian angular velocity. The absorption
coefficient $\kappa_\nu$ is then given by Kirchhoff's Law. In
this work, exactly the same RIAF model parameters in \citet{Y03}
and \citet*{Y06} are adopted to preserve the good fit to the
spectrum of Sgr A* and other observations.

To construct an observed emission structure, we then perform a
full radiation transfer calculation using the ray-tracing method
outlined in Section 2. Here, the radiative transfer equation takes
the form:
\begin{eqnarray}\label{radtrans}
\frac{dI_{\nu}}{ds} = j_{\nu} - \kappa_{\nu}I_{\nu}.
\end{eqnarray}
Since the frequency is red-shifted along the geodesic and
$I_\nu/\nu^3$ is a Lorentz invariant \citep[see, e.g., ][]{Ch06,
Schn06, Nb07}, the observed specific intensity is related to the
emitted intensity by
\begin{eqnarray}\label{Iinv}
I_{\nu_{\rm{obs}}} = \delta^3I_{\nu_{\rm{em}}},
\end{eqnarray}
where the gravitational red-shift factor is calculated to be
\begin{eqnarray}\label{redshf}
\delta =
\frac{\nu_{\rm{obs}}}{\nu_{\rm{em}}}={\sqrt{1-\frac{2M}{r}}}
\end{eqnarray}
for a non-spinning black hole. We neglect the re-emission process
when solving the radiation transfer equation.

Furthermore, light rays emitted from the RIAF bend to us due to
strong gravitation of the black hole, suffer absorption and in
addition, suffer interstellar scattering. That means, what we can
observe is a scatter-broadened image. The two-dimensional
scattering structure in the direction to the Galactic Centre is
determined from fitting to the angular size measurements as a
function of the observing wavelength \citep[cf.][]{Shen05}. It is
a Gaussian ellipse with full-widths at half maximum (FWHM) of
major axis and minor axis in milli-arcsecond (mas) to be
$(1.39\pm0.02)\lambda^2$ and $(0.69\pm0.06)\lambda^2$ ($\lambda$
measured in $\rm{cm}$), respectively and position angle
$\sim80^\circ$ \citep{Shen05}. We take the conversion that one
gravitation radius corresponds to $~5\mu$as angular size for Sgr
A* with $4\times10^6M_\odot$ mass at 8kpc distance.

In Fig. 2, we show both the un-scattered (ray-tracing only) and
scatter-broadened images at wavelengths of 1.3 mm (first two
columns of panels) and 3.5 mm (last two columns), respectively,
obtained from our simulations with parameters $\Theta=0^\circ$,
$i=\pi/2,\pi/4,0$ (from top to bottom). For each two columns at
two wavelengths, right one is obtained by convolving the left one
(GR ray-tracing result) with the interstellar scattering.  We use
color to show the self-normalized specific intensity of the
images.  The abscissa axes point to west and the ordinal axes
point to north. The non-scattered (intrinsic) images are different
at different wavelengths. They exhibit predicted shadows as a
result of relativistic effects.  The scattered image becomes more
obscured as the observing wavelength increases. It indicates that
gradually with the availability of the high-resolution VLBI
imaging at wavelengths of 1.3 mm or shorter, we will be at a very
good position to unveil the shadow structure of the black hole.

\subsection{Comparison with the VLBI Observations at 7
and 3.5 mm}

Currently, VLBI observations can be steadily performed at
millimeter wavelengths of 3.5 and 7 mm. Attempts to determine the
Sgr A* structure with the VLBI observations, however, have
suffered from the angular broadening caused by the diffractive
scattering by the turbulent ionized interstellar medium, which
dominates the resultant images with a $\lambda^2$-dependence
apparent size. The development of the model fitting analysis by
means of the amplitude closure relation along with the careful
design of the observations at millimeter wavelengths has greatly
improved the accuracy of the size measurements of the observed
image \citep*[e.g.,][]{Shen03,Bo04,Shen05}. As a result, an
intrinsic source size as compact as 1 AU was first detected at 3.5
mm \citep{Shen05}. However, for the reason discussed in
\citet{Y06}, it is more meaningful to directly compare the
observed apparent VLBI image with the scatter-broadened image
obtained from the simulations. The apparent images of Sgr A* can
be depicted quantitatively by an elliptical Gaussian distribution
with the following parameters of the major and minor-axis sizes
\citep*{Shen05}: 0.724$\pm$0.001 mas by 0.384$\pm$0.013 mas and
0.21$^{+0.02}_{-0.01}$ mas by 0.13$^{+0.05}_{-0.13}$ mas at
wavelengths of 7 and 3.5 mm, respectively. The position angles at
both wavelengths are about 80$^\circ$, consistent with the
orientation of the scattering structure.

While \citet{Y06} only considered a special configuration, i.e.,
the RIAF is face-on ($i=0^\circ$ and $\Theta=0^\circ$), here we
will do a thorough investigation on the possible geometry of the
RIAF with respect to the presumed supermassive black hole in the
Galactic centre based on the available VLBI measurements at both
3.5 and 7 mm. For this purpose, we first made a series of
simulated after-scattering images with different geometric
configuration (different combination of $i$ and $\Theta$) between
the black hole and the RIAF. Then, we fit the simulated images
with an elliptical Gaussian distribution to estimate the FWHMs of
the major and minor axes.

In Fig. 3, the FWHMs of the major and minor axes of the simulated
scattering-broadened images of Sgr A* are plotted as a function of
the inclination angle ($i$) and the position angle ($\Theta$),
respectively. Thick solid lines are the upper and lower limits of
the FMHMs with the best-fit scattering angular size of
1.39$\lambda^2\times$0.69$\lambda^2$, and thick dashed lines are
the upper and lower limits to account for the $\pm1\sigma$
($0.02\lambda^2\times0.06\lambda^2$) in the scattering size. The
measured angular sizes from high-resolution VLBA imaging
\citep*{Shen05} are indicated in Fig. 3 by the thin solid lines
with the $1\sigma$ uncertainties by the thin dotted lines. As
already shown in \citet{Y06}, within the uncertainties of the
measurements and calculations, the predicted sizes from the RIAF
model with $i=0^\circ$ and $\Theta=0^\circ$ are in reasonable
agreement with the observations at two wavelengths. With a more
thorough consideration shown in Fig. 3, it seems to suggest a
geometry of the RIAF in Sgr A* to have a large inclination angle
$i$ and a small position angle $\Theta$. But it should be cautious
since the results in the minor axis are not accurate, especially
at 3.5 mm \citep*[see,][]{Shen05}. It is clear to us that in order
to get a reliable estimate of the geometry ($i$ and $\Theta$) of
the RIAF in Sgr A*, more accurate measurements are needed.

\section{Visibility Functions of The Black Hole Shadow Images}

As shown in Fig. 2 \citep*[also see Fig. 1 in][]{Y06}, the
simulated apparent scatter-broadened images of Sgr A* at
wavelengths of 1.3 mm and shorter cannot be well fitted by one
elliptical Gaussian component, indicating that the interstellar
medium scattering effects at 1.3 and 0.8 mm no longer dominate the
observed morphology. So, it is very promising to image the shadow
of the supermassive black hole Sgr A* with the high-resolution
VLBI observations at sub-millimeter wavelengths. However, there is
no such a sub-millimeter VLBI array available at present for the
imaging observations. Therefore, we will not focus on the
comparison of the images at wavelengths of 1 mm and shorter, but
try the visibility analysis to show that some useful constraints
on the emission of Sgr A* can be extracted from the limited
visibility data without imaging.

In the interferometric observation, the visibility function
$V(u,v)$ measured on a baseline with coordinates $(u,v)$ is
related, by the Fourier transform, to the sky brightness
distribution, i.e., the shadow image in our simulation. To provide
instructive information to the real observation, we first perform
two-dimensional Fourier transform to obtain the visibility
function of shadows like those in Fig. 2. Then we analyze the
visibility distribution along four specific directions in the sky
plane with position angles of (i) $80^\circ$ (E$^\prime$), i.e.,
along the major axis of the scattering structure; (ii) $-10^\circ$
(N$^\prime$), i.e., along the minor axis of the scattering
structure; (iii) $35^\circ$ (NE$^\prime$); and (iv) $-55^\circ$
(NW$^\prime$). In Fig. 4, we show an example of such visibility
slices with parameters, $i=78.75^\circ$ and $\Theta=-70^\circ$ at
$\lambda=1.3\rm{mm}$. The abscissa is projected baseline length in
the units of $10^9\lambda$ and the ordinate the normalized
visibility amplitude. We plot slices in solid, long-dashed,
short-dashed, and dotted line for E$^\prime$, N$^\prime$,
NE$^\prime$, and NW$^\prime$ directions, respectively. To depict
these visibility slices, we introduce two characteristic baseline
lengths: $\Sigma$ (marked with open circle) to denote the baseline
length at which a normalized visibility decreases to $0.5$, and
$\sigma$ (marked with filled circle) to denote the baseline length
at which the visibility reaches its first valley. Such a minimum
in the visibility distribution is important because it implies
that the image can no longer be described by a single Gaussian but
must have an additional structure, which could be related to the
black hole shadow. Visibilities at baselines longer than $\sigma$
have quite limited signal-to-noise ratio at millimeter
wavelengths. The typical detection limit with the current VLBA
(Very Long Baseline Array) at 3.5 mm is about $0.12$~mJy.

It is obvious that different sets of geometrical parameters ($i$
and $\Theta$) will result in different visibility profiles at
different wavelengths. For a given orientation of the RIAF model
at a given wavelength, we can obtain two characteristic baseline
lengths ($\Sigma$ and $\sigma$) along each of the four directions
(E$^\prime$, N$^\prime$, NE$^\prime$, and NW$^\prime$), i.e.
($\Sigma_{E^\prime}$, $\sigma_{E^\prime}$), ($\Sigma_{N^\prime}$,
$\sigma_{N^\prime}$), ($\Sigma_{NE^\prime}$,
$\sigma_{NE^\prime}$), and ($\Sigma_{NW^\prime}$,
$\sigma_{NW^\prime}$). We further list these in the order of their
lengths and denote them as $\Sigma_{n}$ and $\sigma_{n}$ with n=1
(the shortest) to 4 (the longest).

Our simulations show that the scattering effect at 3.5~mm is still
quite large (see Fig. 2), thus we cannot get enough information of
any fine structure of the shadow. To minimize the scattering
effect, we consider the observations to be performed at other two
shorter wavelengths 1.3 and 0.8~mm. Shown in Fig. 5 are the two
characteristic baseline lengths ($\Sigma$ and $\sigma$) as a
function of the inclination angle $i$ at 1.3 (upper two panels)
and 0.8~mm (lower two panels). With a specific $i$, each of
$\Sigma_n$ and $\sigma_n$ (n=1$-$4) can vary within a region due
to the different position angle $\Theta$. These allowed regions
for $\Sigma_1$, $\Sigma_2$, $\Sigma_3$, and $\Sigma_4$ (or,
$\sigma_1$, $\sigma_2$, $\sigma_3$, and $\sigma_4$) are shown as
the darkest grey with long-dashed border lines, lightest grey with
dotted border lines, second lighter grey with short-dashed border
lines, and white with solid border lines, respectively. From these
plots, we summarize some useful properties as follows:

\begin{enumerate}
\item  Most of the characteristic baseline lengths are in a
    range of (1$-4)\times10^3~\rm{km}$, which should be achievable under
    current VLBI observation conditions.  Therefore, VLBI observations
    of Sgr A* at sub-millimeter wavelengths are expected to provide very
    important results in the near future. Some extreme cases may
    require a project baseline length longer than 5$\times10^3~\rm{km}$ for $\sigma_4$.
\item  When the RIAF is nearly face-on ($i\approx0^\circ$),
    the four $\Sigma$s (or $\sigma$s) are converged to a single similar
    number, suggesting that the four visibility profiles are almost
    identical and thus not orientation-dependent. This is mainly due
    to the isotropy of the intrinsic shadow image without scattering.
    The roughly East-West (80$^\circ$ in position angle) elongated
    scattering structure is very compact, $\le 15 \mu$as along the
    major axis at about 1 mm wavelength, and thus will not affect
    the symmetry of the final scatter-broadened image.
\item With a moderate inclination angle $i$, the differences
    between these projected baselines at different directions are
    quite significant. In fact, the four $\Sigma$s (or $\sigma$s)
    are always range in four separate regions (as shown in Fig. 5)
    because of the asymmetry of the image. The possible region,
    caused by the different position angle $\Theta$, for each $\Sigma$
    and $\sigma$ becomes larger when the inclination angle $i$ increases.
\end{enumerate}

Therefore, in principle we would be able to constrain the
inclination angle ($i$) of the RIAF in Sgr A* by comparing the
above-mentioned characteristic baseline lengths determined from
the future 1.3 and/or 0.8 mm VLBI experiments with the simulation
demonstrated in this paper. It can be inferred from Fig. 5 that
when the RIAF has a large inclination angle (in our simulation of
Sgr A*, $i>\sim 30^\circ$), both $\Sigma$s and $\sigma$s become
very sensitive to $\Theta$.

No wonder, the position angle $\Theta$ should have effects on the
characteristic baseline lengths too. For clarity, we define here
another three normalized differences in the characteristic
baseline lengths:

\begin{eqnarray}\label{ss}
S_1 &=&
(\Sigma_{N^\prime}-\Sigma_{E^\prime})/(\Sigma_{4}-\Sigma_{1}), \nonumber\\
S_2 &=&
(\Sigma_{NE^\prime}-\Sigma_{NW^\prime})/(\Sigma_{4}-\Sigma_{1}), \\
S_3 &=&
(\Sigma_{E^\prime}-\Sigma_{NE^\prime})/(\Sigma_{4}-\Sigma_{1}).\nonumber
\end{eqnarray}
Note that $\Sigma_{4}$ and $\Sigma_{1}$ represent the longest and
shortest characteristic baseline lengths, respectively. If the
RIAF is nearly face-on ($i\approx0^\circ$), the scatter-broadened
image of Sgr A* is roughly E-W elongated, resulting in the
$\Sigma_{N^\prime}$ being the longest and $\Sigma_{E^\prime}$ the
shortest (i.e., $\Sigma_{N^\prime}$=$\Sigma_{4}$ and
$\Sigma_{E^\prime}$=$\Sigma_{1}$), and
$\Sigma_{NE^\prime}$=$\Sigma_{NW^\prime}$ because of its symmetry.
Therefore, it is always the case that $S_1=1$, $S_2=0$ and
$-1<S_3<0$ for any $\Theta$. This means the face-on case is not
sensitive to the position angle, consistent with the almost
identical characteristic baseline lengths shown in Fig. 5 when
$i\approx0^\circ$.

Shown in Fig. 6 are plots of $S_1$, $S_2$ and $S_3$ as a function
of the position angle ($\Theta$). At a given position angle
$\Theta$, these values vary with the inclination angle $i$ and
have a range represented by grey region with dashed border lines
(at 1.3 mm) and dotted region with dotted border lines (at
0.8~mm).

Assuming that the RIAF model is a physically reasonable
description of the observed emission of Sgr A*, our simulation
indicates that starting with the limited visibility measurements
on the four baselines in the directions mentioned above from the
future sub-millimeter (1.3 and 0.8 mm) VLBI experiments, we have a
very good chance to get tight constraints on the configuration
($i$ and $\Theta$) of the RIAF around the central supermassive
black hole of Sgr A*. That is, we can understand the geometry of
the radio-emitting region surrounding Sgr A* in advance, even
without imaging it with a completely $(u,v)$ coverage of
sub-millimeter/millimeter VLBI array. It is important to indicate
that the detection of $\sigma$s, i.e., valleys in the visibility
profile (see Fig. 4 and Fig. 5) , are very important to prove that
there is indeed a supermassive black hole in the centre of our
Galaxy.

\section{Discussion and Summary}
With the emissivity and absorption coefficient from the
radiatively inefficient accretion flow model for the Galactic
Centre black hole candidate Sgr A*, we solved the radiative
transfer using the ray-tracing code to get the simulated shadow
images of Sgr A* with an arbitrary set of geometrical parameters,
i.e., $i$ and $\Theta$. By taking into account the interstellar
scattering, we can obtain the apparent images at different
wavelengths.

Comparison of the simulated image sizes with those measurements at
both 7 and 3.5 mm seems to prefer a RIAF with a large inclination
angle and small position angle for Sgr A*. But, large
uncertainties in both the size measurements and the interstellar
scattering relationship make these very uncertain at the moment.
Even though, our simulations show that the future sub-millimeter
VLBI images of Sgr A* would reveal a significant deviation from
the single Gaussian structure, which can be used to probe the
geometry of the RIAF of Sgr A*.

Analysis of the scatter-broadened images also show that an
observing wavelength of 1.3 mm or shorter is needed to identify
the shadow shape. However, currently it is difficult to construct
a VLBI array operated at 1.3 mm to obtain such an image. We
therefore first perform the Fourier transform to obtain the
corresponding visibility functions of images at 1.3 and 0.8 mm,
then analyze these visibility profiles along some specific
directions. By introducing several characteristic baseline lengths
($\Sigma_n$ and $\sigma_n$) and the normalized differences
($S_n$), we demonstrate that it is possible to determine
fundamental parameters ($i$ and $\Theta$) of Sgr A* from the
limited detections on some baselines of VLBI observations at a
wavelength of 1~mm or shorter (cf. Fig. 5 and Fig. 6). We can
understand the geometry of the radio-emitting region surrounding
Sgr A* even without imaging it with a completely $(u,v)$ coverage
of a sub-millimeter/millimeter VLBI array.

It should be mentioned that the four directions chosen in our
analysis are somehow arbitrary. Any other combinations would also
provide useful constraints on the structure of Sgr A*. For
example, it would be interesting to choose such two specific
slices, one along the galactic plane and the other perpendicular
to it. In practice, we can perform the similar simulations with
the projected baselines to be exactly the same as those from the
real sub-millimeter VLBI experiments in the future. Of
course the RIAF model adopted here is still quite simple.  We set
the static black hole, use Pseudo-Newtonian potential \citep{Y03},
and simply assume a Gaussian distribution of the density as the
vertical structure of the accretion flow. And the possible
dynamical role of an ordered magnetic field is not considered. We
also neglect the Doppler effect of the radial and angular motions
of the accretion flow in the radiative transfer calculation.
However in principal, the visibility analysis outlined in this
paper should be applicable, even if Sgr~A* is a spinning black
hole or with other accretion flow geometries and emission models.
As a necessary first step trying to solve this complicated
problem, this work definitely needs further efforts in both
modelling and observations. And a more quantitative study will be
required in future work.

Furthermore, that the characteristic baseline lengths are about
$10^3$~km and the expected observing wavelength is about 1 mm
encourage us to propose the development of sub-millimeter VLBI
array.  The image of the vicinity of the predicted black hole in
Sgr A* is expected to be more easily detected than any other
super-massive black hole candidates because of its closest
distance to us. If the shadow image could be finally observed, it
will be strong evidence for the General Relativity theory.

\section*{Acknowledgments}

We would like to thank two anonymous referees for their critical
comments on the manuscript. This work was supported in part by the
National Natural Science Foundation of China (grants 10573029,
10625314, and 10633010) and the Knowledge Innovation Program of
the Chinese Academy of Sciences, and sponsored by Program of
Shanghai Subject Chief Scientist (06XD14024). Z.-Q. Shen and F.
Yuan acknowledge the support by the One-Hundred-Talent Program of
Chinese Academy of Sciences.

\begin{figure*}
    \includegraphics[width=0.85\textwidth]{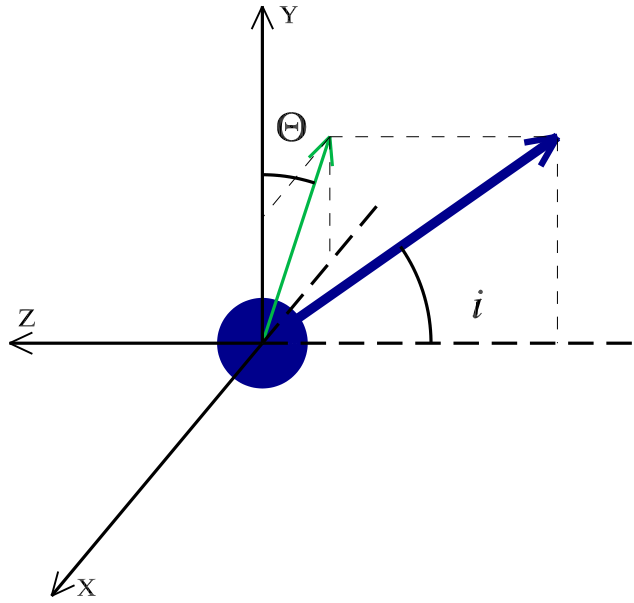}
    \caption{Sketch map of the black hole.  $X$-axis is to east, $Y$-axis to north and $Z$-axis
    aligns with the line of sight of an observer.
    The blue arrow represents the direction of the black hole's equatorial plane which is aligned
    with the mid-plane of the RIAF (see text), and $i$ and $\Theta$ are the inclination angle
    and position angle, respectively.}
\end{figure*}

\begin{figure*}
  \includegraphics[width=0.85\textwidth]{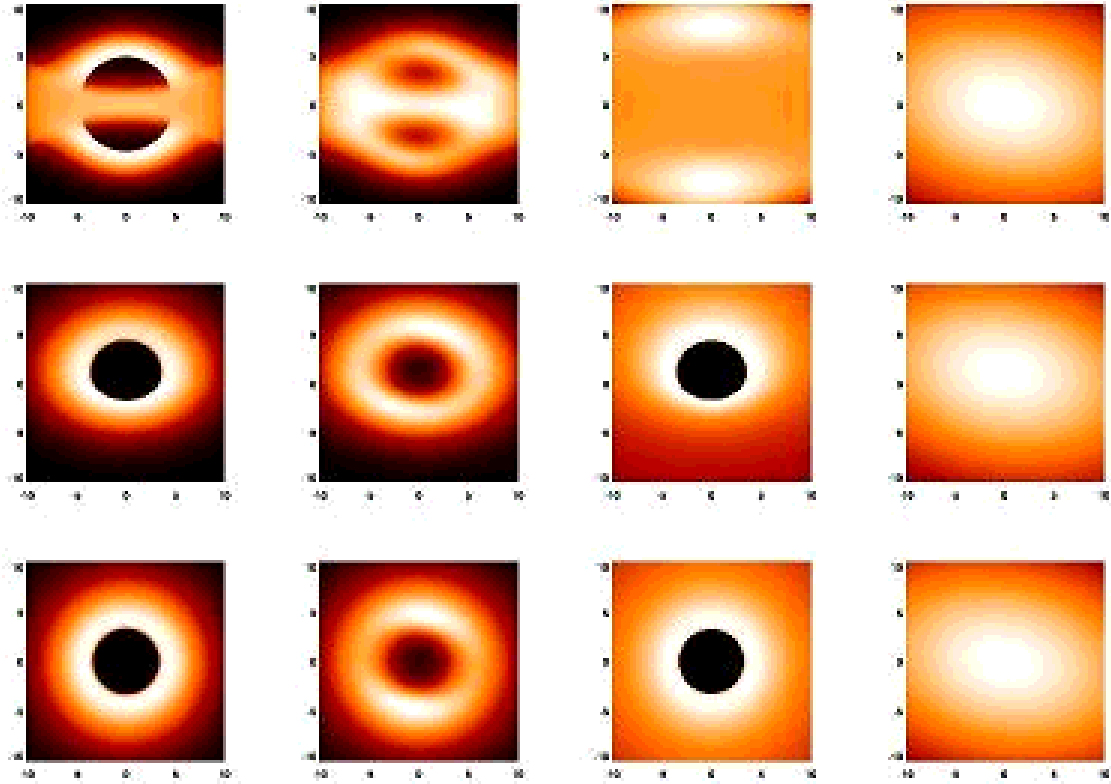}
  \caption{Simulated images of the RIAF of Sgr A* with
$\Theta=0^\circ$ and $i=\pi/2 ,\pi/4,0$ (from top to bottom) at
wavelengths of 1.3 mm (left two columns of panels) and 3.5 mm
(right two columns of panels). For each two columns at two
wavelengths, right one is obtained by convolving the left one (GR
ray-tracing result) with the interstellar scattering.  In each
panel, the intensity is normalized to itself to show detailed
structure. The units in the X- and Y-axes are gravitational radius
$r_g$, which is 6$\times10^{11}$ cm or 5$\mu$as for Sgr A* with
$4\times10^6M_\odot$ mass at 8 kpc distance.}
\end{figure*}

\begin{figure*}
  \includegraphics[width=0.4\textwidth]{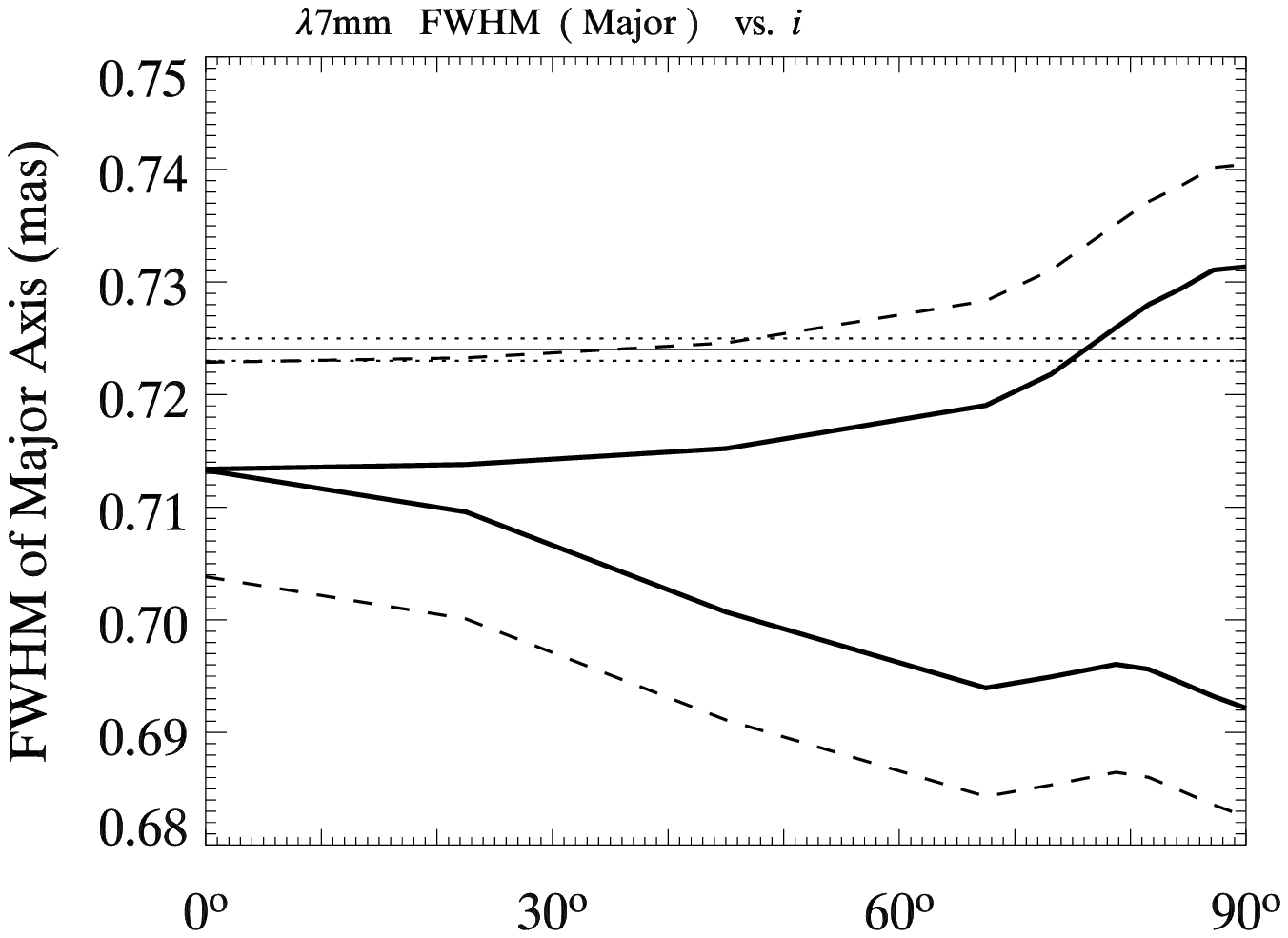}
  \includegraphics[width=0.4\textwidth]{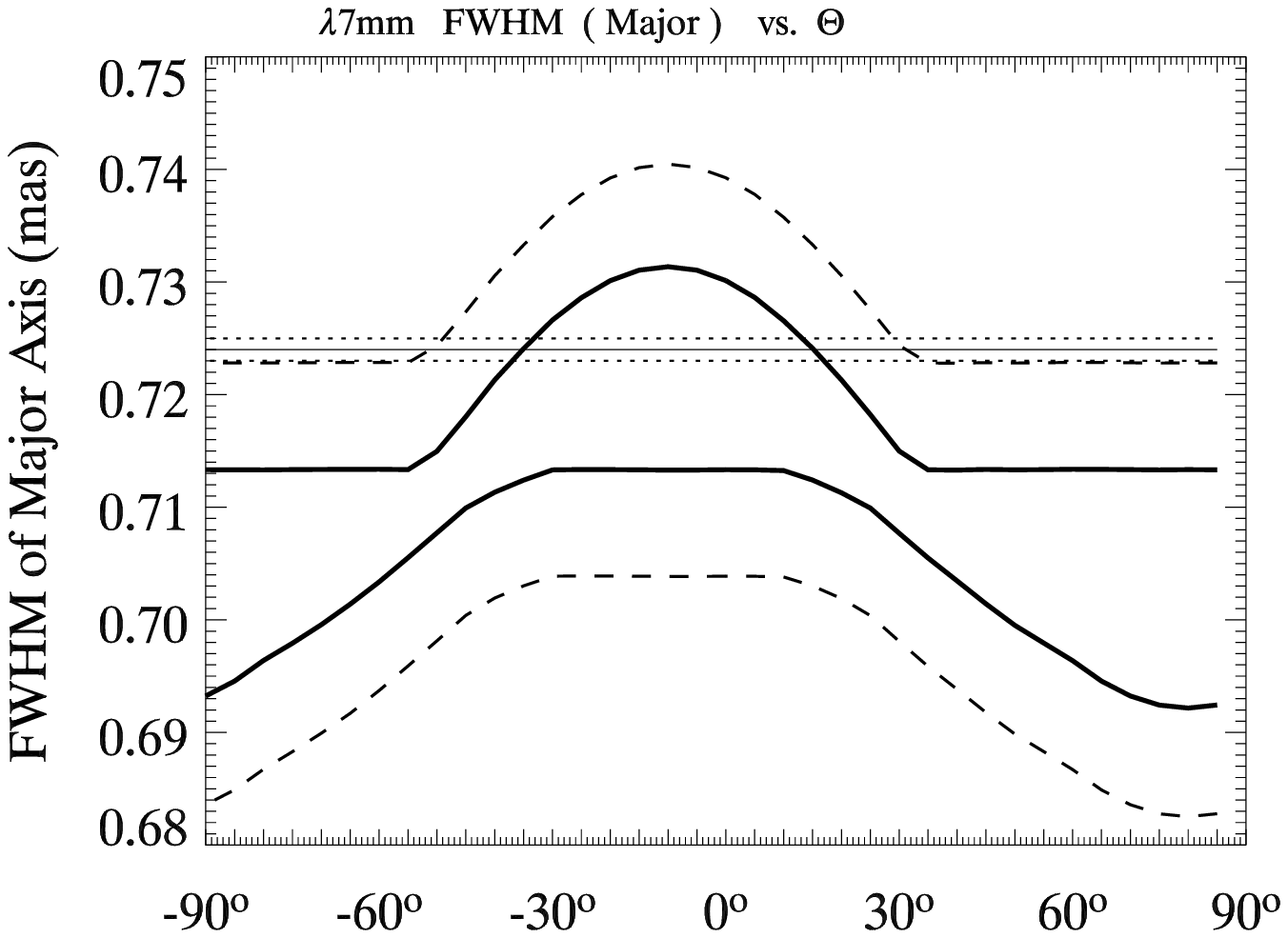}
  \\
  \includegraphics[width=0.4\textwidth]{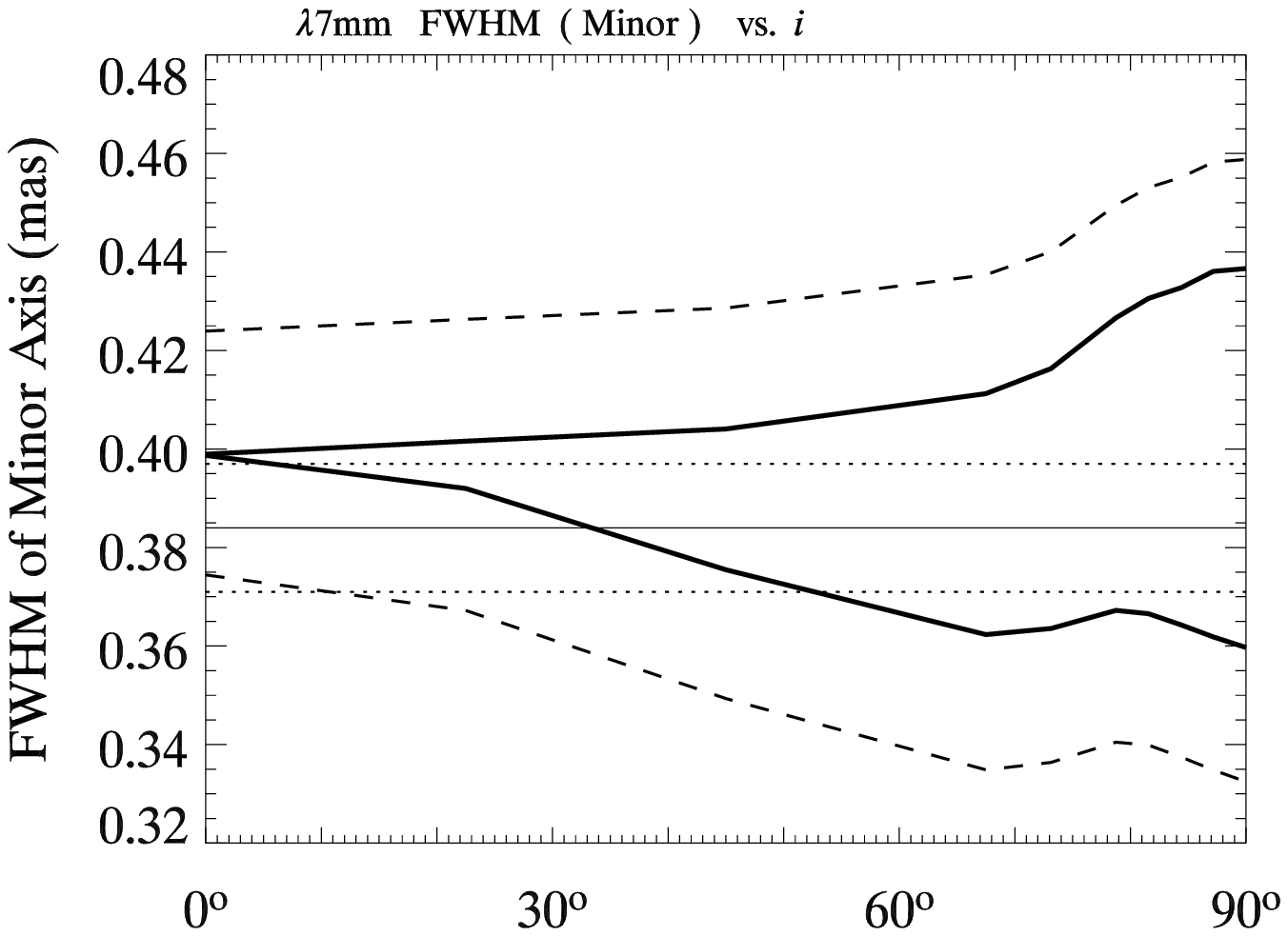}
  \includegraphics[width=0.4\textwidth]{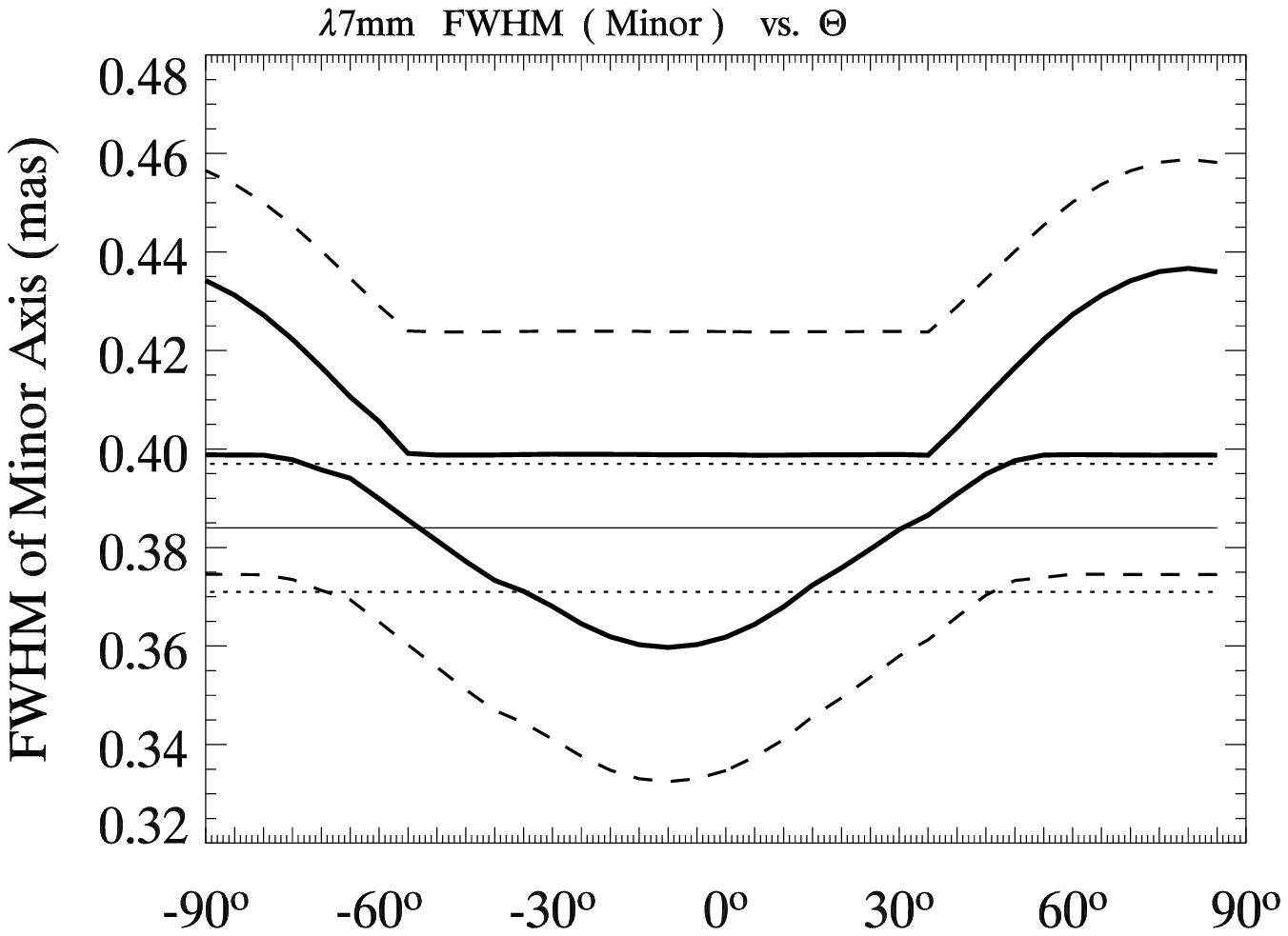}
  \\
  \includegraphics[width=0.4\textwidth]{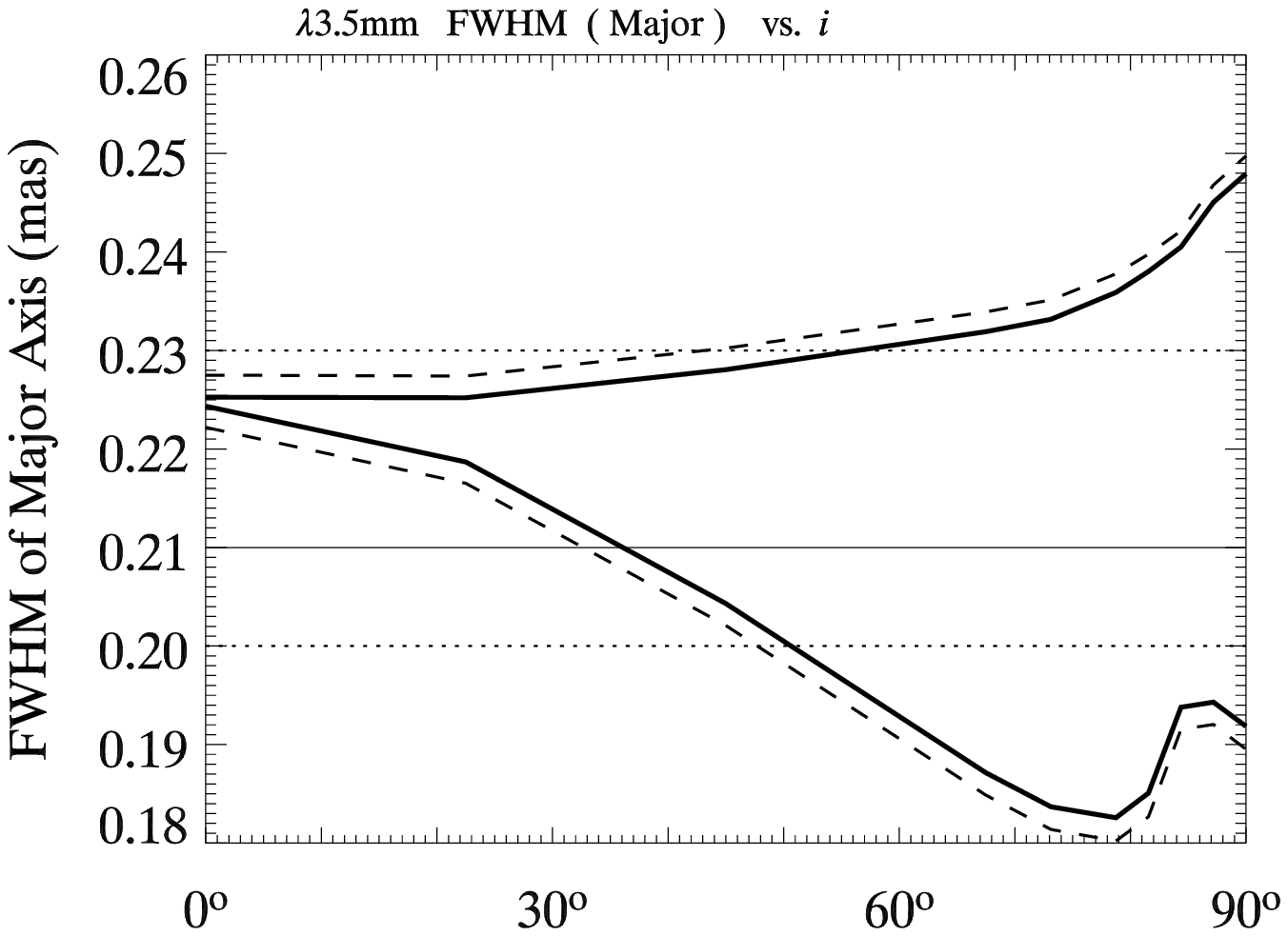}
  \includegraphics[width=0.4\textwidth]{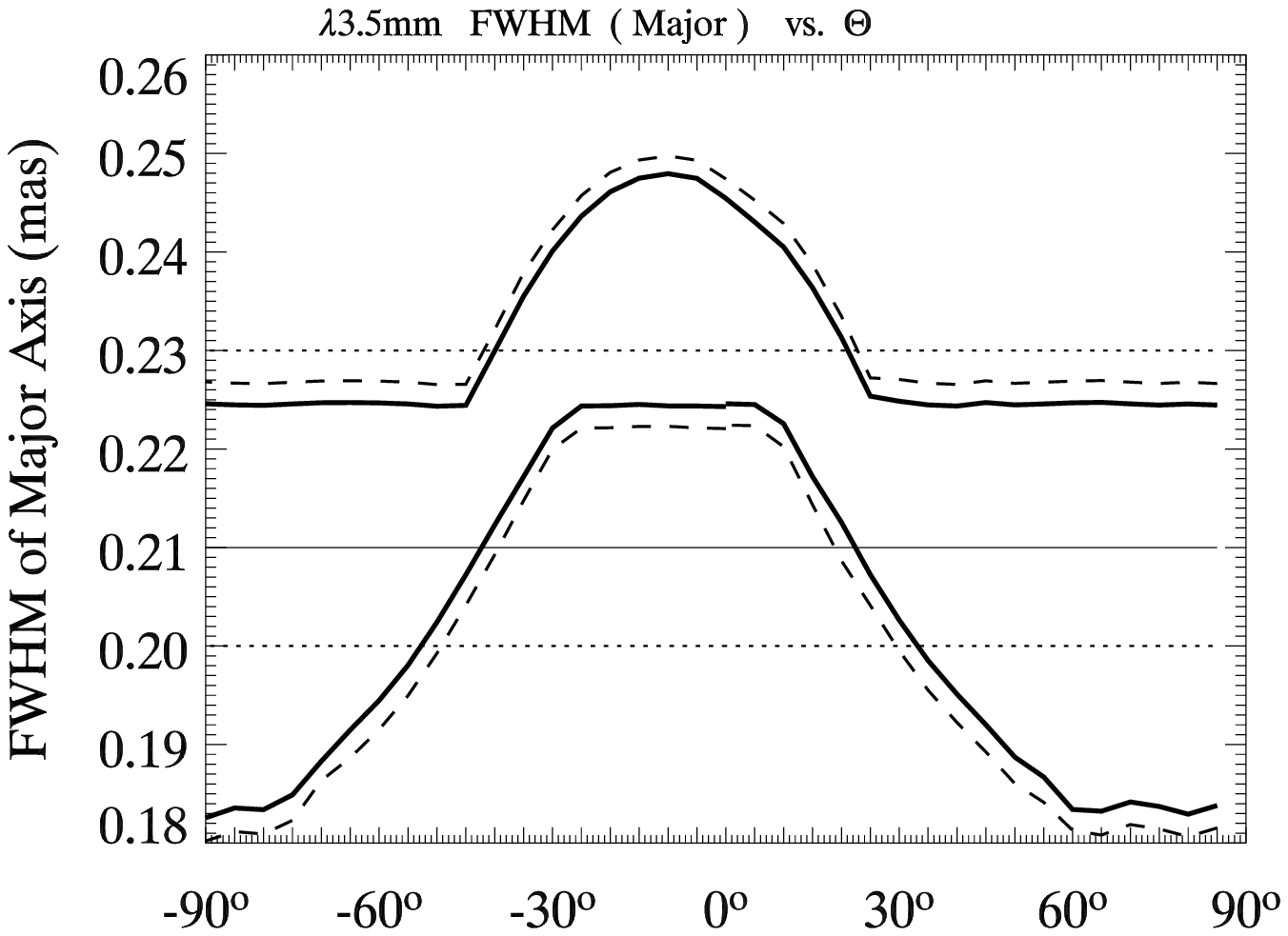}
  \\
  \includegraphics[width=0.4\textwidth]{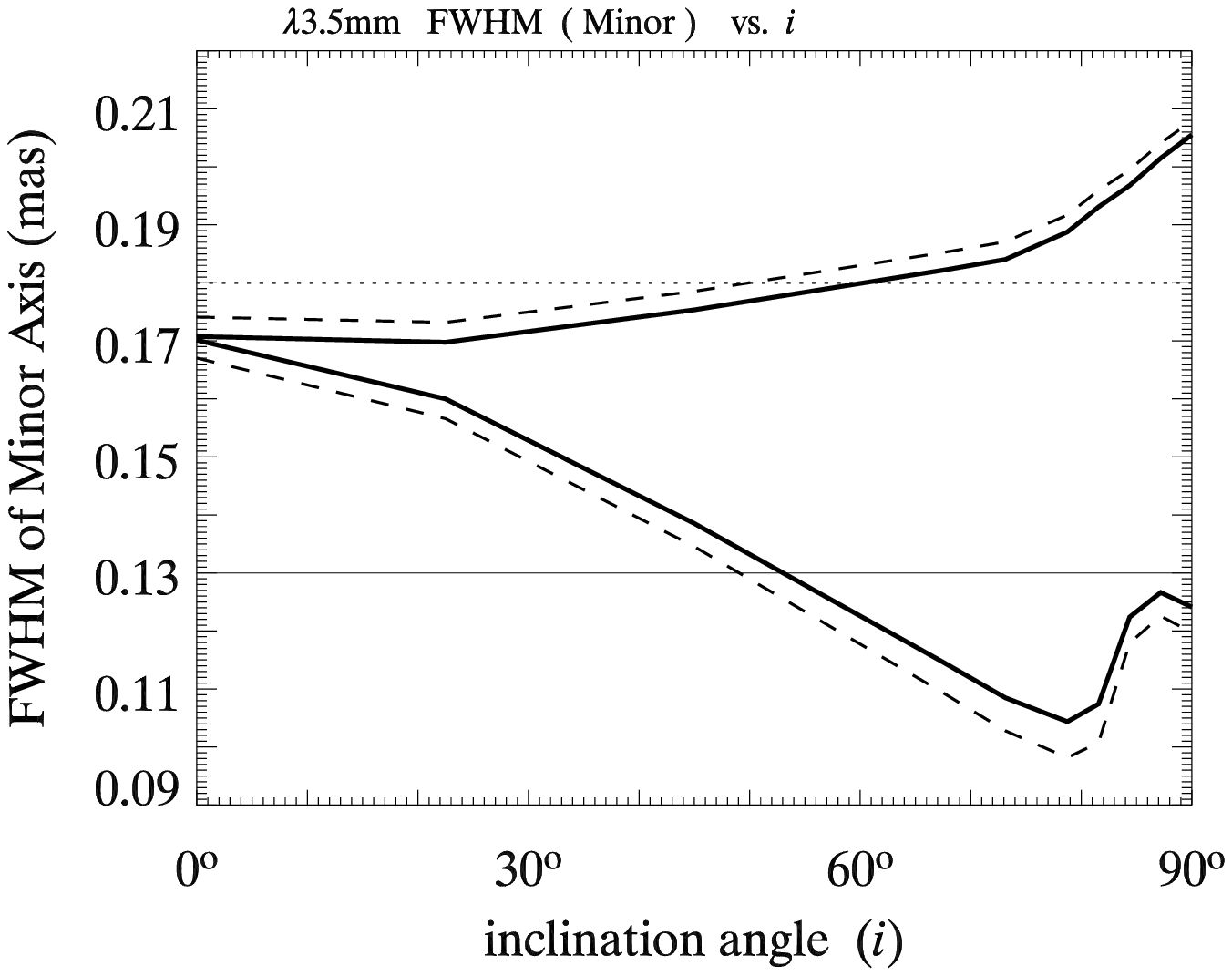}
  \includegraphics[width=0.4\textwidth]{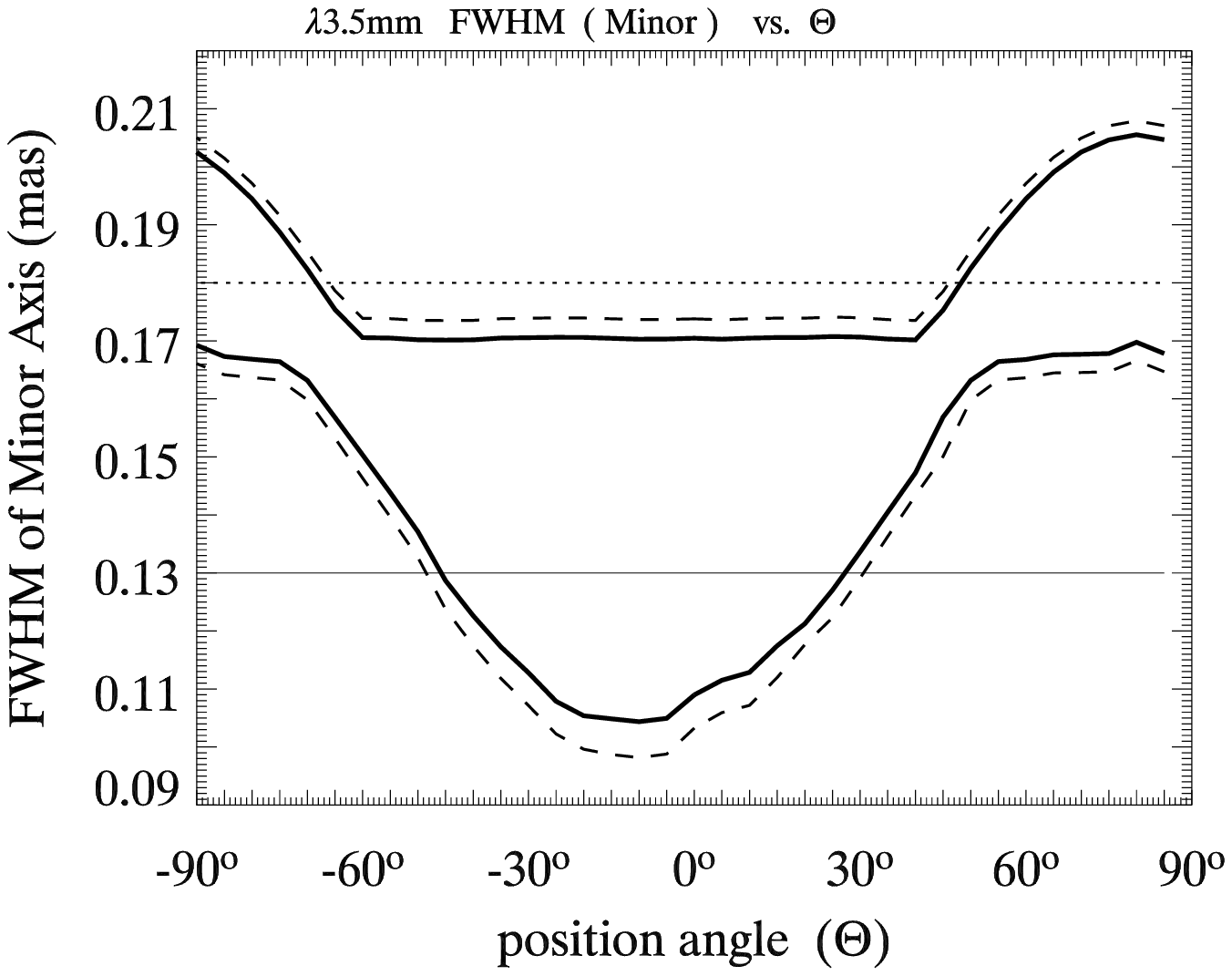}
  \\
\caption{Comparison of the simulated sizes of Sgr A* with the real
measurements at 7 mm (upper four panels) and 3.5 mm (lower four
panels). Panels in left column, from top to bottom, show sizes
(i.e. FWHMs) of the scatter-broadened images along directions of
major axis at 7 mm, minor axis at 7 mm, major axis at 3.5 mm, and
minor axis at 3.5 mm, as a function of the inclination angle ($i$)
of the RIAF, while the panels in right column show these sizes as
a function of the position angle ($\Theta$) of the RIAF. In each
panel, the thick solid lines represent upper and lower limits of
FWHMs of the images broadened by a constant scattering effect of
1.39$\lambda^2\times$0.69$\lambda^2$ with a position angle of
80$^\circ$. The upper and lower limits to the sizes are due to
different geometrical structure, i.e. different position angles at
a fixed inclination angle $i$, or different inclination angles at
a fixed position angle $\Theta$. After considering the
uncertainties in the scattering angular sizes, the upper and lower
limits are plotted with the thick dashed lines. The measured sizes
and errors are represented by those straight thin solid and dotted
lines.}
\end{figure*}

\begin{figure*}
  \includegraphics[width=1.0\textwidth]{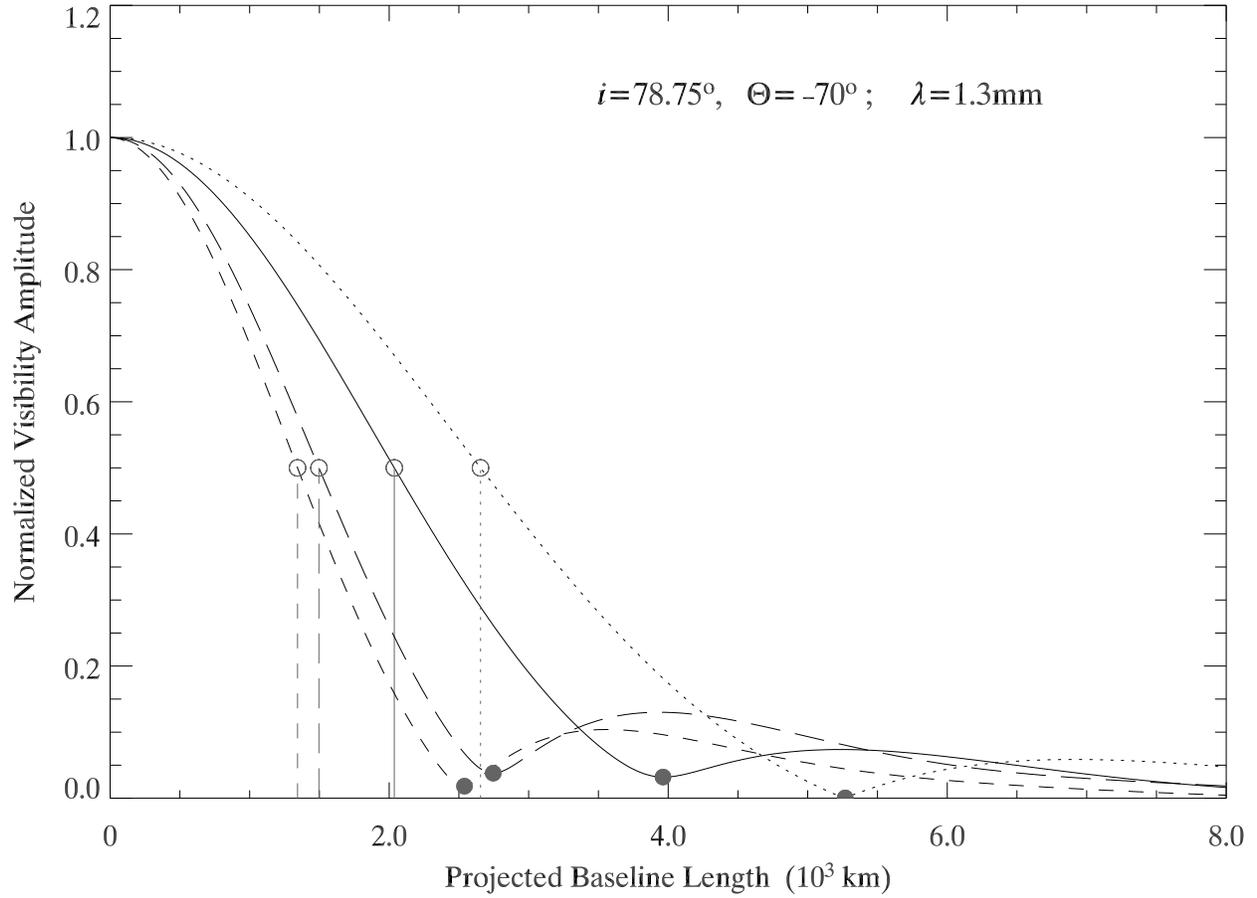}
\caption{The predicted visibility function of Sgr A* (in this
case, at 1.3 mm, with $i$=78.75$^\circ$ and $\Theta$=
$-70^\circ$). The four curves represent the normalized visibility
amplitude profiles at four position angles, i.e. 80$^\circ$ (solid
line), $-10^\circ$ (long-dashed line), $35^\circ$ (short-dashed
line) and $-55^\circ$ (dotted line). We mark where a profile
reaches 0.5 (open circle) and its first valley (filled circle).}
\end{figure*}

\begin{figure*}
  \includegraphics[width=0.45\textwidth]{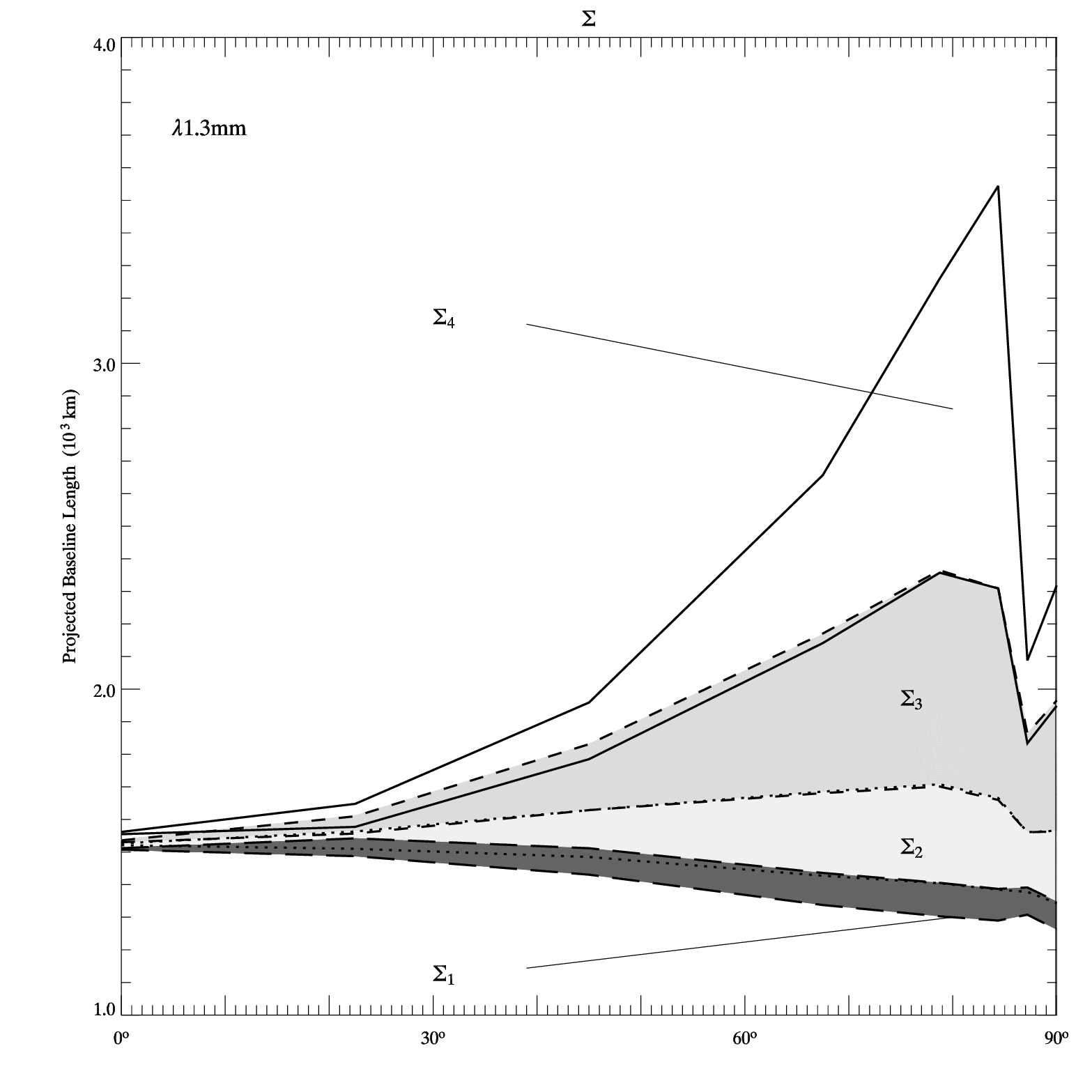}
  \includegraphics[width=0.45\textwidth]{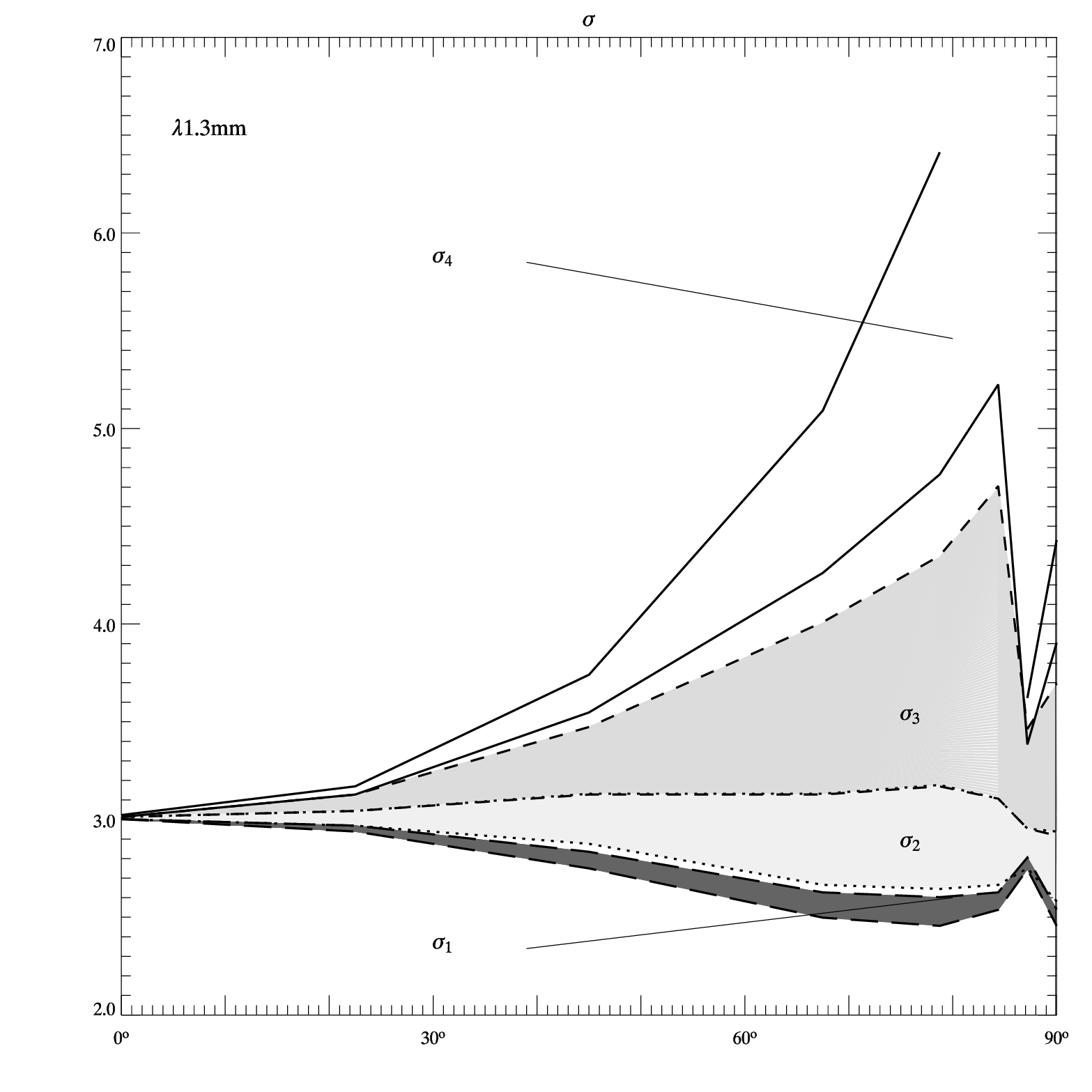}
  \\
  \includegraphics[width=0.45\textwidth]{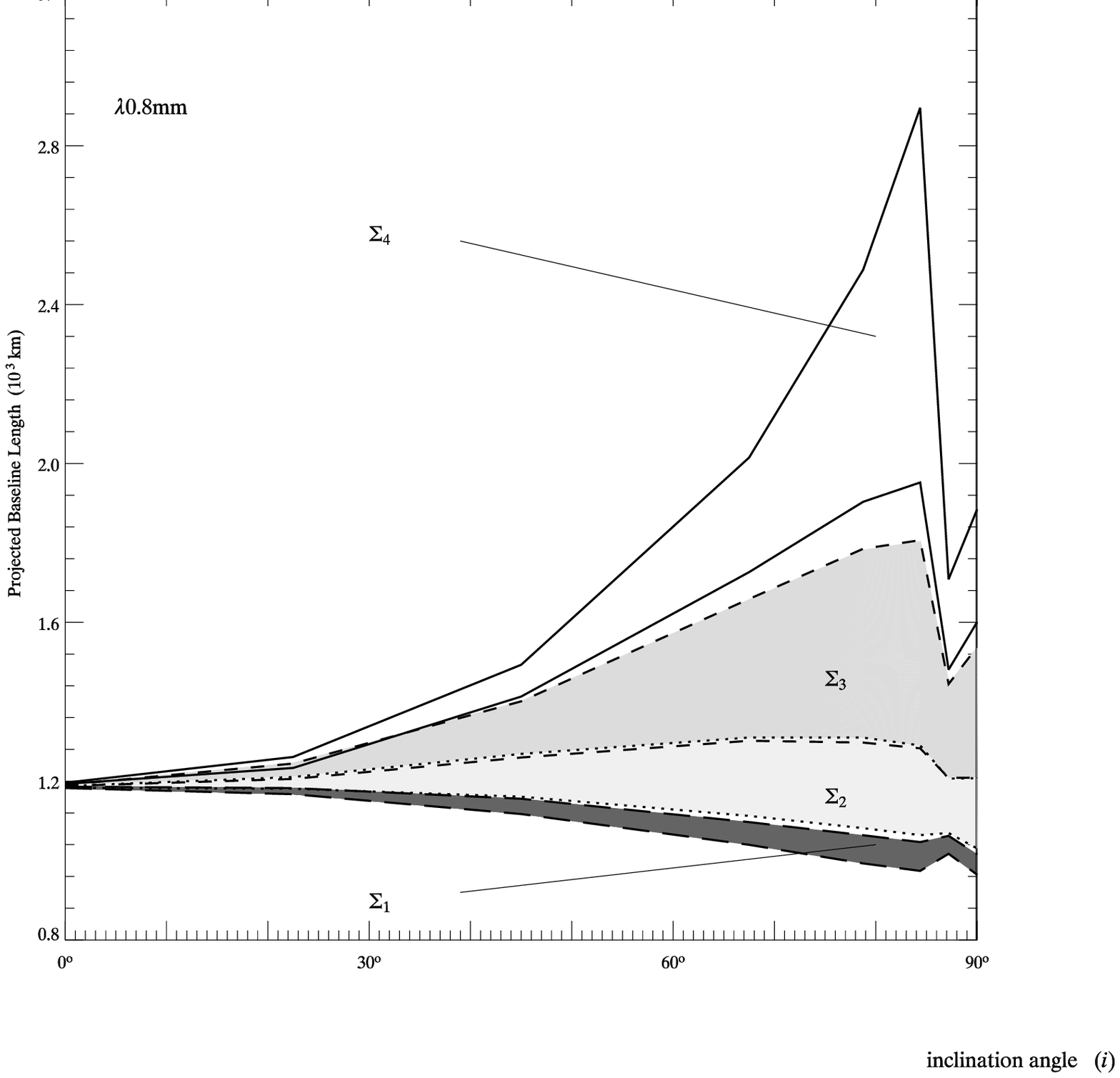}
  \includegraphics[width=0.45\textwidth]{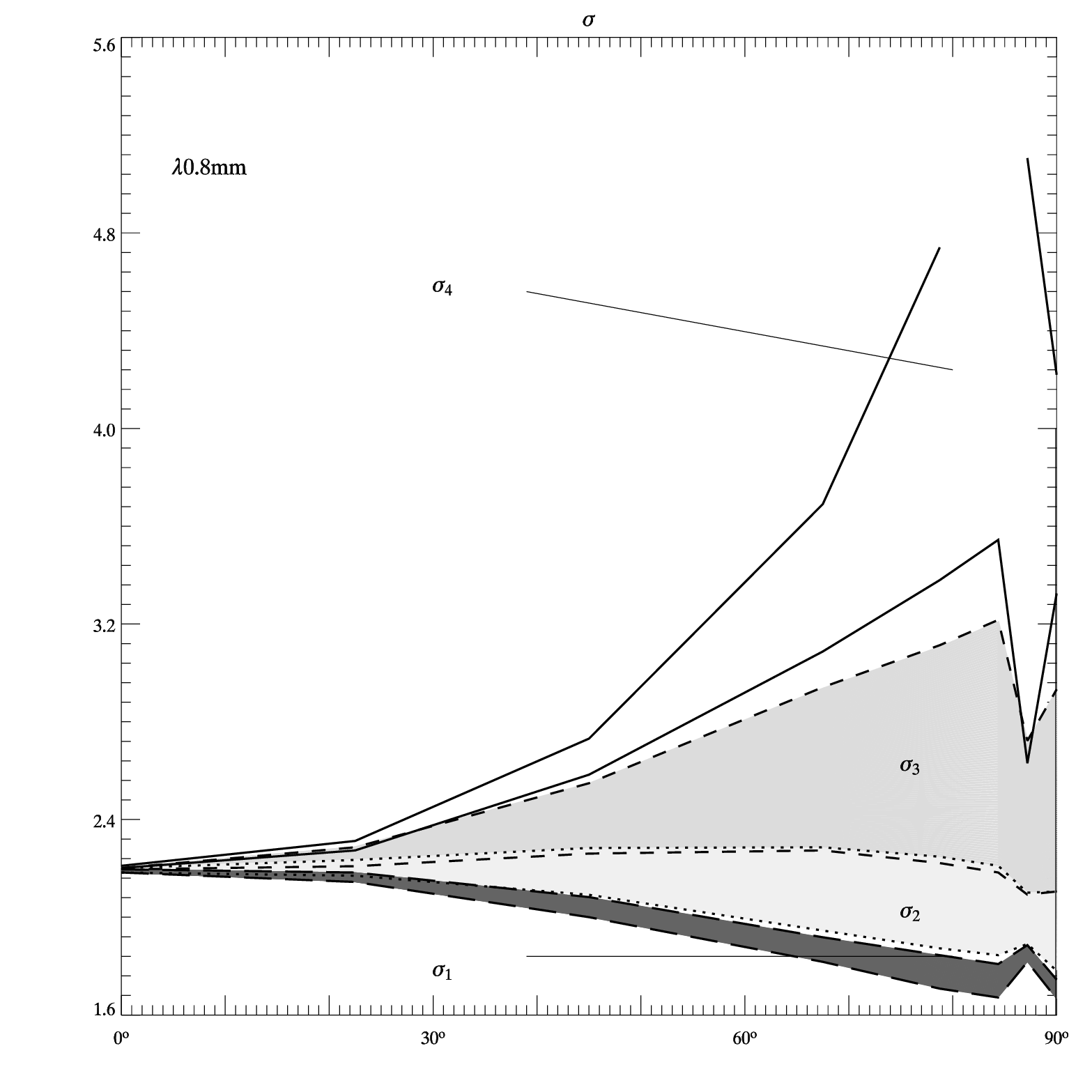}
  \\
\caption{The projected characteristic baseline lengths versus the
inclination angle $i$ at 1.3 mm (upper two panels) and 0.8 mm
(lower two panels). With a specific inclination angle $i$, each of
$\Sigma_n$ and $\sigma_n$ (n=1-4) can vary with the position angle
$\Theta$. These allowed regions for $\Sigma_1$, $\Sigma_2$,
$\Sigma_3$, and $\Sigma_4$ (or, $\sigma_1$, $\sigma_2$,
$\sigma_3$, and $\sigma_4$) are shown as the darkest grey with
long-dashed border lines, lightest grey with dotted border lines,
second lighter grey with short-dashed border lines, and white with
solid border lines, respectively.}
\end{figure*}

\begin{figure*}
  \includegraphics[width=1.0\textwidth]{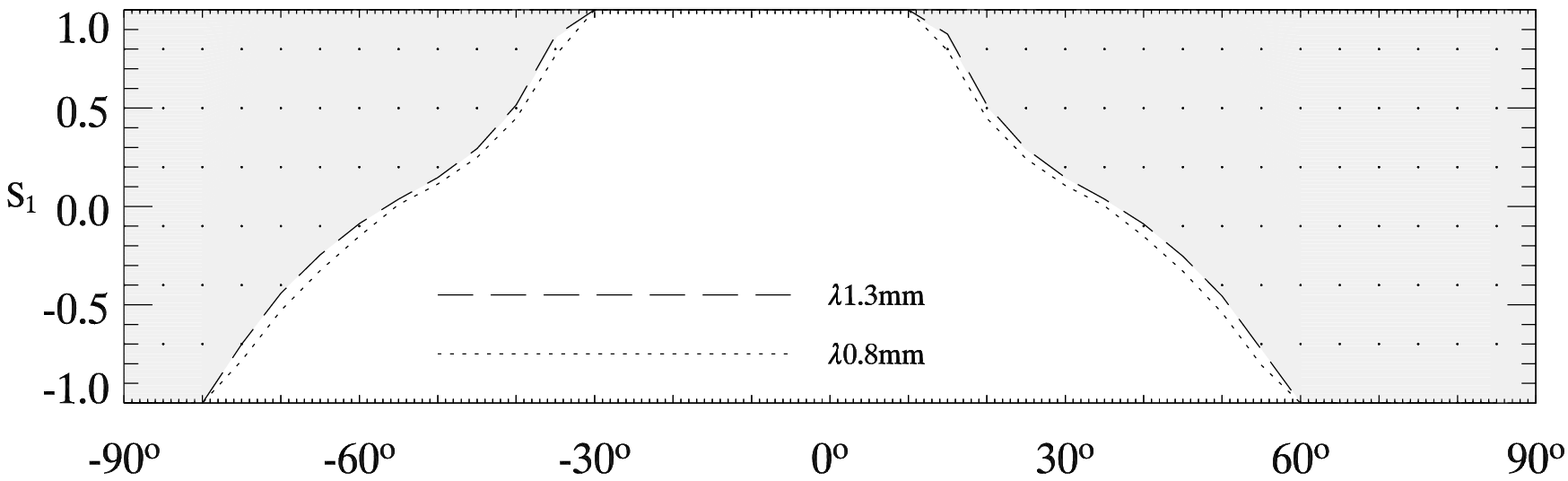}
  \\
  \includegraphics[width=1.0\textwidth]{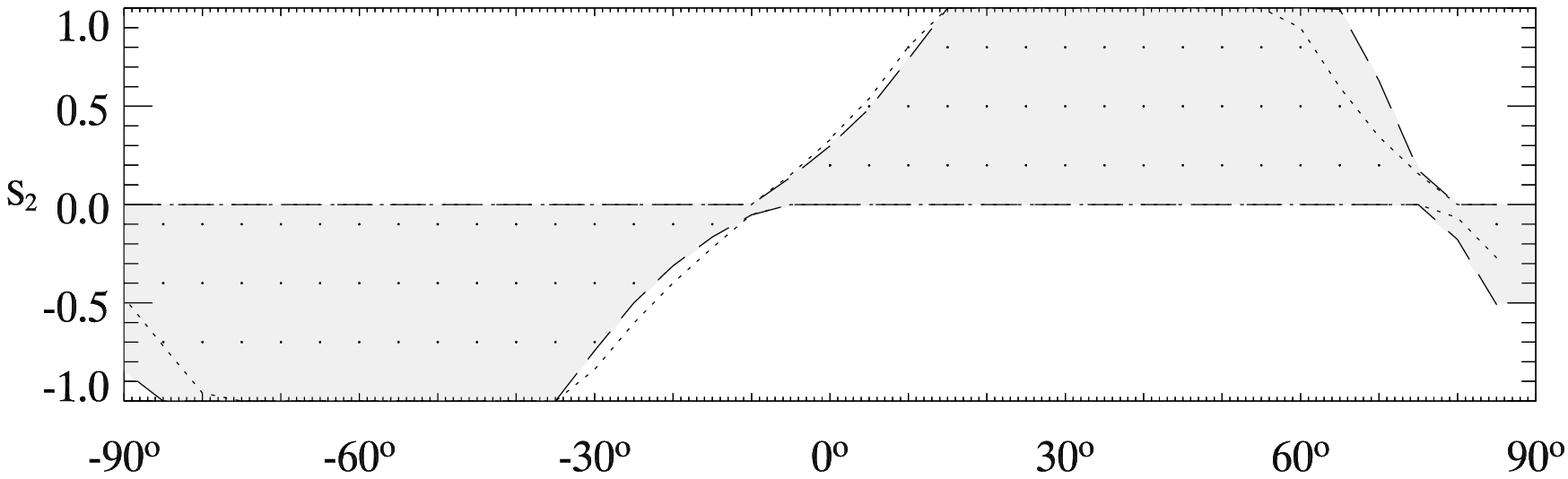}
  \\
  \includegraphics[width=1.0\textwidth]{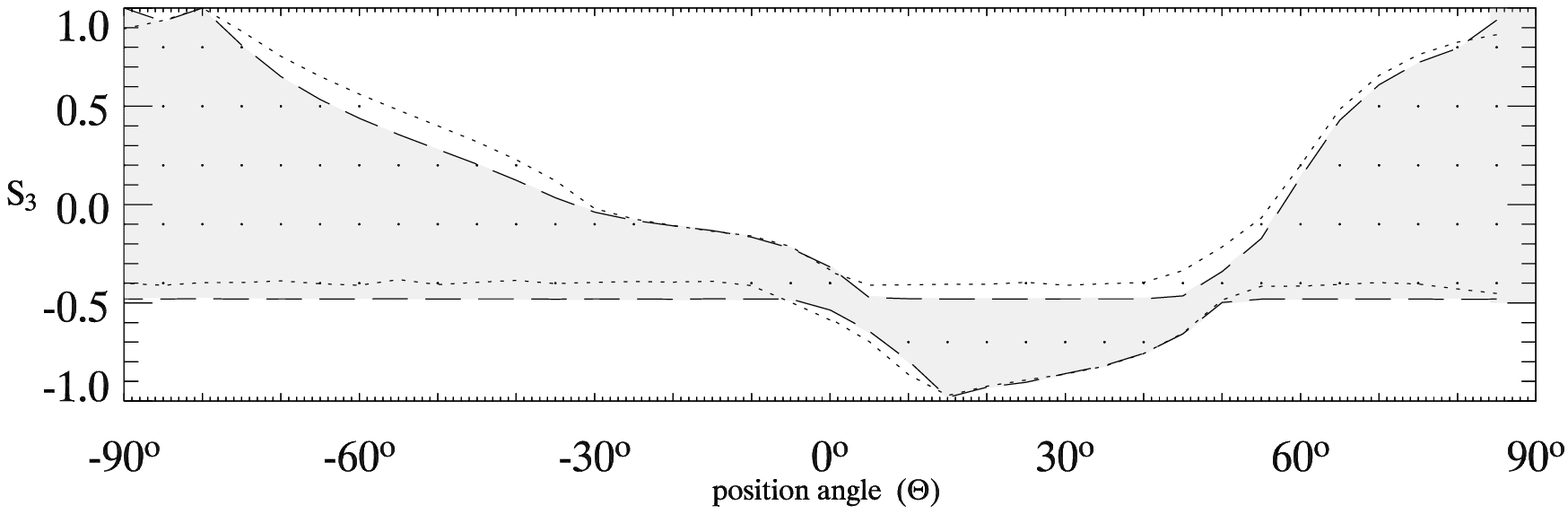}
  \\
\caption{The normalized difference of the characteristic baseline
lengths ($S_1$, $S_2$, and $S_3$, from top to bottom) as a
function of the position angle $\Theta$. For a given position
angle ($\Theta$), $S_1$, $S_2$, and $S_3$ can vary with the
inclination angle $i$, and have a range represented by the grey
region with dashed border lines and the dotted region with dotted
border lines for observations at 1.3 and 0.8 mm, respectively.}
\end{figure*}

\bsp

\label{lastpage}

\end{document}